\documentclass[11pt]{article}
\usepackage{hyperref}
\usepackage{amsmath}
\usepackage{cite}
\usepackage{subfloat}
\usepackage{graphicx}
\usepackage{relsize}
\usepackage{amssymb}
\usepackage{float}
\usepackage[bottom=1 in,top=1.3 in]{geometry}

\begin{document}
\title{\textbf{Effective interaction in a non-Fermi liquid conductor and spin correlations in under-doped cuprates}}
\author{\textbf{Suraka Bhattacharjee$^*$} and \textbf{Ranjan Chaudhury$^\dagger$} \\ 
Department of Condensed Matter Physics and Material Sciences\\S.N. Bose National Centre for Basic Sciences\\Saltlake, Sector-III, Block-JD, Kolkata-700106\\
Email- surakabhatta@bose.res.in$^*$, ranjan@bose.res.in$^\dagger$}
\maketitle
\vspace*{0.5cm}
\section*{Abstract}
The effective interaction between the itinerant spin degrees of freedom in the paramagnetic phases of hole doped quantum Heisenberg antiferromagnets is investigated theoretically, based on the single-band t-J model on 1D lattice, at zero temperature. The effective spin-spin interaction for this model in the strong correlation limit, is studied in terms of the generalized spin stiffness constant as a function of doping concentration. The plot of this generalized spin stiffness constant against doping shows a very high value of stiffness in the vicinity of zero doping and a very sharp fall with increase in doping concentration, signifying the rapid decay of original coupling of semi-localized spins in the system. Quite interestingly, this plot also shows a maximum occurring at a finite value of doping, which strongly suggests the tendency of the itinerant spins to couple again in the unconventional paramagnetic phase. As the doping is further increased, this new coupling is also suppressed and the spin response becomes analogous to almost Pauli-like. The last two predictions of ours are quite novel and may be directly tested by independent experiments and computational techniques in future. Our results in general receive good support from other theoretical works and experimental results extracted from the chains of YBa$_2$Cu$_3$O$_{6+x}$. \\

\section{Introduction}
\hspace*{0.3cm} 
The complete understanding of the microscopic physics for low-dimensional bulk systems in the strongly correlated domain still remains a very challenging problem in spite of the growing interest in nano and mesoscopic physics. Some of the strongly correlated doped quantum antiferromagnets are also well known to exhibit high temperature superconductivity in the vicinity of optimal doping concentration \cite{1,2}. The unconventional itinerant character of these doped antiferromagnets can be substantially explored within the framework of the strongly correlated t-J model. In the earlier works, the nature of the effective interaction between the static holes in doped La$_{2-x}$Sr$_x$CuO$_4$ was analytically studied in the very low doping region (x$\sim$0.05) \cite{3}. The result showed that the addition of holes introduces ferromagnetic coupling between the Cu$^{++}$ spins and the frustration leads to the emergence of the spin glass state in the under-doped regime \cite{3}. Recently, the phase diagram of the Hubbard model and frustrated Hubbard model has been studied using the density matrix embedding theory \cite{4}. The strong coupling diagram technique also revealed the influence of the long range spin and charge fluctuations in the long range antiferromagnetically ordered phase of the two-dimensional Fermionic Hubbard model, with large repulsions and at very low temperature \cite{5}. The above technique showed the destruction of this antiferromagnetic ordering with the increase in doping concentration \cite{5}. \\
\hspace*{0.3cm} The phase separation in the 2D t-J model was studied by minimizing the total energy with respect to the number of holes in the hole rich and hole deficient phase \cite{6}. The holes were only considered as the “missing spins” on a particular site, neglecting their actual spin configuration \cite{6}. A more rigorous quantum mechanical approach was proposed in our earlier paper to study the effective interaction between the spin and charge degrees of freedom of the carriers separately in the conducting phase of a strongly correlated system in 2D. This would help in identifying the magnetic correlations as well as pairing possibilities. In 1D strongly correlated conducting phase, this is even more relevant indeed and very important as there are exotic possibilities like spin-charge decoupling \cite{7}. There were many former attempts to study the magnetic correlations in the doped layered systems using the Mori's projection technique based on two-time thermodynamic Green's function and Variational Monte Carlo simulations \cite{8,9,10,11}. In a previous paper of ours, we have determined the effective magnetic couplings and predicted the various possible magnetic phases of the doped quantum antiferromagnets on a two-dimensional (2D) lattice. In our calculation we had imposed the strict `no double occupancy condition'(NDOC) on each site in the ground state wave function, characterizing a non-Fermi liquid (NFL) state \cite{12}. We had studied the interactions between the itinerant spin degrees of freedom in details, in the under-doped phase, in particular with our calculation of generalized spin stiffness constant ($\tilde{D_s}$) for this model \cite{12}. Moreover, we could predict a point of possible quantum phase transition in the over-doped regime, with our strongly correlated model extended to that region \cite{12}. Our calculational results for the generalized spin stiffness constant corresponding to a single pair of mobile holes (D$_s$) were compared with the experimental observations on layers of La$_{2-x}$Sr$_x$CuO$_4$ \cite{12,13}. The comparison established the role of D$_s$ as an effective intra-layer exchange constant between a pair of holes in the quasi-2D doped antiferromagnetic materials, at least in the regime of low doping concentration ($\delta$) \cite{12}. \\
\hspace*{0.3cm}Inspired by the successful application of our formalism in 2D, doped quantum antiferromagnetic models, in the present paper we have undertaken the theoretical investigation of the doping dependence of effective spin-spin coupling for the strongly correlated t-J model on the 1D lattice. The aim of this study is to explore the spin dynamics of this model as well and find possible application to chained cuprates in the underdoped phases. The results for 1D model are found to be in sharp contrast to our results for that of 2D, implying once again the strong lattice-dimensional dependence of low-dimensional magnetism. \\
\hspace*{0.3cm}Various other theoretical approaches were used earlier for investigating the magnetic properties of the t-J model and they are briefly discussed below. The different phases of the 1-D t-J model was studied using the exact diagonalization method \cite{14}. The presence of a phase separated state was confirmed by the divergence of thermodynamic compressibility around the critical value of the ratio of J/t lying between 2.5 to 3.5, for various electron densities \cite{14}. But the above results could not predict anything about the interactions between the existing spins \cite{14}. The 1D t-J model is exactly solvable using the Bethe Ansatz for the specific parameter ratio of J/t=0 and J/t=2 only \cite{15}. The correlation function for this model was also derived only for the special case of J=t. Thus for general t/J ratio this technique can not be utilized. In this context, the Density Matrix Renormalization Group (DMRG) and Transfer Matrix Renormalization Group (TMRG) came up as attempts to study the spin correlations away from the super-symmetric points of the t-J model\cite{16,17}. The outcomes show a general trend of fall in singlet correlation, resulting in an increase of spin susceptibility as the electron number density is decreased \cite{17}. The above numerical calculations were done at certain chemical potentials, which accounts for the dependence of spin correlation on electron densities; however, no estimation of the exact values doping concentrations and doping dependence of exchange coupling could be given \cite{17}. Thus there is a genuine need for an alternative approach to be attempted. This provides us with a very strong motivation to employ our scheme based on $\tilde{D_s}$ \cite{12}. This would further give us a solid ground for detailed comparison of our calculational results obtained for the 1D and 2D t-J models. The comparison of the results is expected to bring out the essential microscopic physics behind the significant differences in the features of the spin correlations observed in various quasi-1D and quasi-2D doped antiferromagnetic real materials.

In 2D our calculated exchange energy contribution to spin stiffness constant (D$_s^J$) shows a drastic fall with increase in $\delta$ in the low doping region, whereas, the kinetic energy contribution (D$_s^t$) remains zero throughout the low doping regime \cite{12} This is in very good agreement with the previous experimental and theoretical (Monte Carlo) results \cite{11,13,18}. On the other hand, our calculated D$_s^J$ in 1D falls even more rapidly initially with $\delta$, and then in striking contrast to the 2D case, the D$_s^t$ in 1D shows an increase with $\delta$ in the very low $\delta$ region, followed by a drastic fall again throughout the rest of the doping regime. This behaviour results in the appearance of a new peak in the very low $\delta$ region of the D$^t_s$ vs. $\delta$ plot in 1D.  \\
\hspace*{0.3cm} Our approach and formalism are further enriched by the possible comparison of our derived results in 1D model with the available experimental results from the chains of YBa$_2$Cu$_3$O$_{6+x}$ (x being the doping concentration), which behaves as an  itinerant paramagnet, believed to be describable by the strongly correlated t-J model. \\
\hspace{0.3cm}The role of Cu1-O1 planes in high temperature superconductivity has been a debated issue for a long time, but the necessity of the chains have been established by many researchers and resonant elastic x-ray scattering has identified distinct ordering in chains and planes of YBa$_2$Cu$_3$O$_{6+x}$ \cite{19,20,21,22,23,24}. Initially the oxygen holes enter the chains of YBa$_2$Cu$_3$O$_{6+x}$ followed by entry onto the Cu2-O2 planes beyond a critical value of x (x$_c$) \cite{19,20,21,22}. This indicates that the bulk susceptibility measurements on the sample in lower doping regimes corresponds mostly to the response from the chains of the compound \cite{25}. \\
\hspace{0.3cm}Static spin susceptibility of exchange coupled paramagnetic systems is generally calculated from zero frequency limit of experimentally extracted imaginary part of dynamic spin susceptibility (DSS), using the Kramers Kronig relations \cite{26}. Again the static limit of the well known Fluctuation Dissipation Theorem relates the real part of DSS to the static correlation function between the spin degrees of freedom in the system \cite{26}.\\
\hspace{0.3cm} The effective intra-chain wave-vector dependent magnetic exchange constants (J$_{eff}$(q)) can be expressed as the inverse of the wave-vector dependent static spin susceptibility ($\chi$(q)) by using a standard theoretical approach developed for the itinerant magnets \cite{27}. Following this, we have extracted J$_{eff}$(0) i.e. the effective ferromagnetic exchange constant from the experimentally observed uniform dc magnetic susceptibility ($\chi$(0)) of YBa$_2$Cu$_3$O$_{6+x}$ \cite{25}. Again, the scans corresponding to q=Q$\equiv\pi$/a obtained from the inelastic neutron scattering experimental results for the above material are used to extract the effective exchange constant at the antiferromagnetic wave vector Q=$\pi$/a \cite{28}. \\ 
\hspace*{0.3cm} Finally, following our approach, we could handle the short-ranged ordered paramagnetic phases of the strongly correlated doped antiferromagnetic systems in a more comprehensive way, which has been a major challenge for the theoreticians so far \cite{12}. The earlier attempts to study the exchange interactions between the spins could determine the correlations between the localized spins or the itinerant spins only in the weakly correlated regime \cite{27,29,30}. However, the detailed behaviour of exchange interactions between the spins in the strongly correlated itinerant regime remains unexplored till today . In this background, our quantum mechanical approach devised an efficient formalism for finding the exchange constant of strongly correlated semi-itinerant systems, without using the Density Functional approach and Multiple Scattering Theory \cite{26,28,30}.   \\    

\section{{Mathematical Formulation and Calculation}}
The t-J model Hamiltonian for strongly correlated electronic systems is \cite{12,32}:
\begin{align}
                   H_{t-J}=H_t+H_J
\end{align}
where H$_t$ and H$_J$ are the kinetic (due to doping) and exchange Hamiltonians respectively for the nearest-neighbour processes. Here it may be recalled that this kinetic energy part arises from the hopping of holes in the doped phase and the exchange part represents the exchange interactions between the itinerant spin degrees of freedom. Moreover, the t-J model gets reduced to the well known localized Heisenberg model corresponding to the completely half-filled band (undoped phase). 
\begin{align}
        H_J=\sum_{<ij>}J_{ij}(\overrightarrow{S_i}.\overrightarrow{S_j}-\frac{1}{4}n_i n_j)
\end{align}
where S$_i$ and S$_j$ now represent the localized spin operators corresponding to the i$^{th}$ and j$^{th}$ sites respectively; J$_{ij}$ is the exchange constant involving the i$^{th}$ and the j$^{th}$ site and for nearest neighbour pair $\langle$ij$\rangle$, J$_{ij}$=J; n$_i$ and n$_j$ are the occupation number operators for the i$^{th}$ and j$^{th}$ site respectively.
\begin{align}                                  
                 H_t=\sum_{<i,j>,\sigma}t_{ij}X^{\sigma0}_iX^{0\sigma}_j
\end{align}
Here t$_{ij}$ represents the hopping amplitude from j$^{th}$ to i$^{th}$ site and for nearest neighbour t$_{ij}$=t. The X's are the Hubbard operators that satisfy the Hubbard algebra and the commutation relation:
\begin{align}
             [X_i^{\alpha\beta},X_j^{\gamma\delta}]=\delta_{ij}(\delta^{\beta\gamma}X_i^{\alpha\delta}-\delta^{\alpha\delta}X_i^{\gamma\beta})
\end{align}
To avoid the rather complicated algebra of the Hubbard operators, for simplicity in our calculation we have used the Fermion operators satisfying the usual anti-commutation relation and have utilized the relations between the spin and the Hubbard operators \cite{12,32}:
\begin{align}
     S_+=X^{+-}\hspace{1cm} S_-=X^{-+}\hspace{1cm}S_z=\frac{1}{2}(X^{++}-X^{--})
\end{align}
where the symbols used for all the spin operators have their usual meanings and they represent the itinerant spin operators. \\
 The generalized spin stiffness constant ($\tilde{D_s}$) can be expressed as \cite{12,32}:
 \begin{align}
 \tilde{D_s}=\tilde{D_s^t}+\tilde{D_s^J}
\end{align}  
where $\tilde{D_s^t}$ and $\tilde{D_s^J}$ are the contributions to spin stiffness constant from kinetic energy and exchange energy 
respectively and are given by \cite{12,32}:
     \begin{align}
    \tilde{D_s^t}=\lim_{\phi\rightarrow0}(\frac{1}{2})\frac{\delta^2T}{\delta\phi^2}
\end{align} and
\begin{align}
    \tilde{D_s^J}=\lim_{\phi\rightarrow0}(\frac{1}{2})\frac{\delta^2E_J^{sf}}{\delta\phi^2}
\end{align}
where $\phi$ is the magnetic twist corresponding to the staggered Peierl's phase $\phi_\sigma$ arising from the presence of the vector potential A($\overrightarrow{r}$), with the property \cite{12,32}:
 \begin{align}
 \phi_\downarrow=-\phi_\uparrow=\phi
\end{align}
 with `T' being the expectation value of the kinetic energy part of the Hamiltonian (1) and `E$_J^{sf}$' is the spin flip contribution to the expectation value of exchange energy part of the Hamiltonian \cite{12,32}.\\
 The hopping amplitude `t' gets modified to t$_{ij}e^{i\phi_\sigma}$ with the inclusion of the Peierl's phase $\phi_\sigma$, if A($\overrightarrow{r}$) has a component along the direction of hopping \cite{32}.\\
 The energy expectation values are calculated in the Gutzwiller state, the proposed variational ground state, with the double occupancy exclusion condition on each site \cite{33}:
 \begin{align}
 \vert\psi_G\rangle=\prod_l(1-\alpha\widehat{n}_{l\uparrow}
 \widehat{n}_{l\downarrow})\vert{FS}\rangle
 \end{align}
 where $\vert{FS}\rangle$ is the non-interacting Fermi sea and the variational parameter $\alpha$ denotes the amplitude for the projection out of the doubly occupied sites corresponding to the strongly correlated systems. For very strongly correlated systems ie.,for infinitely large value of onsite Coulomb repulsion U with respect to bandwidth, the detailed numerical results show that the variational parameter $\alpha\rightarrow$ 1 in the half-filled to low doping regime for the 2D systems \cite{34}. 
As an approximation, we have taken $\alpha$=1 even for our 1D model, implying complete projecting out of the doubly occupied sites \cite{12,32}.\\
 Further, the Fermi sea in equation (10) can be expressed in terms of Fermion creation operators and thus equation (10) becomes: 
 \begin{align}
 \vert\psi_G\rangle=\prod_l(1-\widehat{n}_{l\uparrow}
 \widehat{n}_{l\downarrow})\prod_{k\sigma}^{k_F}\sum_{ij}
 C_{i\sigma}^{\dagger} C_{j-\sigma}^{\dagger}
 e^{i(\overrightarrow{r_i}-\overrightarrow{r_j}).\overrightarrow{k}}\vert{vac}\rangle
\end{align} 
where $\vert{vac}\rangle$, i, j and l have the usual meaning as described in Ref.(10); k is the wave vector bounded  by the Fermi wave vector k$_F$ which is defined with respect to the non-interacting free carriers in the ideal Fermi sea after introduction of vacancies \cite{12}. It might be noted that k$_F$ being the Fermi wave vector for the non-interacting carriers at T=0, all the k-states below k$_F$ are completely filled, as are occurring in the above equation (11).      \\
For 1-D systems, the Fermi wave vector is related to the number of occupied sites as:
\begin{align}
       k_F=n(\pi/2a)
\end{align}
where `a' is the lattice constant and `n' is the fraction of occupied sites in the system defined by:
     \begin{align}
     n=N_l/N=(1-\delta)
     \end{align}
Here `$\delta$' is the doping concentration; `N$_l$' and `N' are the number of occupied lattice sites and the total number of lattice sites respectively. \\
Making use of equations (12) and (13):
       \begin{align}
       k_F=(\pi/2a)(1-\delta)
\end{align} 
Again combining equations (2) and (11), E$_J^{sf}$ can be  expressed as \cite{12}: 
\begin{align}
E_J^{sf}=(\frac{2t_{eff}^2cos(2\phi)}{V_{eff}})
\frac{_{NDOC}\langle\psi_G\vert{H_J'}\vert\psi_G\rangle_{NDOC}}{_{NDOC}\langle\psi_G\vert\psi_G\rangle_{NDOC}}
\end{align}
where 
\begin{align}
H_J'=\overrightarrow{S_i}.\overrightarrow{S_j}-\frac{1}{4}n_in_j
\end{align}with
$_{NDOC}\langle\psi_G\vert\psi_G\rangle_{NDOC}$ being the factor for the normalization of the Gutzwiller state \cite{12}; t$_{eff}$ and V$_{eff}$ are the effective nearest-neighbour hopping and on-site Coulomb barrier potential respectively in the $\delta\rightarrow$0 limit \cite{12,32}. In the case of one-dimensional systems, the initial J in the $\delta\rightarrow$0 limit is modeled as 2t$_{eff}^2/V_{eff}$. Here it must be kept in mind that the investigation for the variation of effective exchange constant with doping concentration has been done by keeping the initial t$_{eff}$ and V$_{eff}$ constant i.e. bare `t' and `J' as constants. \\
 \hspace*{0.3cm}We have carried out detailed rigorous calculations for determining the expectation value of exchange energy in the Gutwiller state. Then taking derivative twice in the $\phi\rightarrow$0 limit, we derive the expression for $\tilde{D^J_s}$ for one-dimension as (for detailed scheme of application see Appendix A of Ref.(10)):
  \begin{align}
  \tilde{D^J_s}=-4J\prod_{k,\sigma}^{k_F}2(1-\delta)^2
\end{align}  
where `J' is the bare exchange constant, as explained earlier. This equation (17) looks very similar to the one for $\tilde{D^J_s}$ as was obtained for 2D, but here k$_F$ corresponds to the Fermi wave vector for 1-D as given in equation (12). It is also seen here that the magnitude of 
 $\tilde{D^J_s}$ analytically goes to zero only for $\delta\rightarrow$1 ie., for 100$\%$ doping concentration, which of course signifies non existence of carriers (holes) in the system!

  Similarly, we have derived the expression for $\tilde{D^t_s}$ (for derivational scheme see Appendix B of Ref.(10)):
  \begin{align}
  \tilde{D^t_s}=(-t)[\prod_{k,\sigma}^{k_F}4cos(ka)(1-\delta)^2-N_l\prod_{k,\sigma}^{k_F}4cos(ka)/N^2]
\end{align}   
The second term in equation (18) appears due to complete projection of the doubly occupied sites and becomes negligible for very high values of $\delta$, as the chances of double occupancy decrease with increase in vacancies in the system \cite{12}.\\
Then from equation (18) one sees that the quantity $\tilde{D^t_s}$ vanishes for $\delta\rightarrow$1. Furthermore, $\tilde{D^t_s}$ also vanishes if at least one value of `k' in the whole set of values of k in the range 0$\leq\vert{k}\vert\leq{k_F}$ is $\pi/2a$ \cite{12,32}. This condition will be satisfied if the upper boundary of k is greater than or equal to $\pi$/2a, which can be ensured with (k$_F)_{threshold}$=$\pi$/2a \cite{12,32}. \\
Then from equation (14),we have
\begin{align}
(\pi/2a)(1-\delta)=\pi/2a 
\end{align}
This condition can only be satisfied for $\delta\rightarrow$0 ie., in the undoped phase.\\
Hence, the spin stiffness constant at the $\delta\rightarrow$0 limit is solely due to the contribution from the exchange energy part and the total $\tilde{D_s}$ vanishes theoretically only for 100$\%$ doping concentration. \\
As was explained in our earlier paper, $\tilde{D_s^J}$ and $\tilde{D_s^t}$ are further scaled down by $^{N_{l}}$C$_2$ which is the number of possible pairs of mobile holes in the system \cite{12}.  \\
 Thus,
\begin{align}
  D_s=\tilde{D_s}/^{N_{l}}C_2
\end{align}
Further, we define the continuation to the generalized spin stiffness from the exchange energy, per pair (D$^J_s$) as \cite{12}:
 \begin{align}
 D^J_s=\tilde{D^J_s}/^{N_{l}}C_2
 \end{align}
 and 
the continuation to the generalized spin stiffness from the kinetic energy, per pair (D$^t_s$) as \cite{12}: 
 \begin{align}
 D^t_s=\tilde{D^t_s}/^{N_{l}}C_2
 \end{align}
 
We have evaluated and plotted D$^J_s$ and D$^t_s$ against $\delta$ for three different lattice lengths viz.(1900,1940,1960) which are presented in the next section. The comparison with other theoretical and experimental results on doped YBa$_2$Cu$_3$O$_{6+x}$ are also presented in the next section of our paper. 
\vspace*{3cm}
 
\section{Our Numerical Results and Comparison with  Experimental and other Computational Results}
We searched for the relevant experimental results on quasi-1D strongly-correlated doped antiferromagnetic systems \cite{35,36,37,38,39,40,41}. The very few available experimental results on the doped SrCu$_2$O$_3$, Sr$_2$Cu$_3$O$_5$, CaCu$_2$O$_3$ etc. could not be used for detailed comparison with our theoretical results on strongly correlated t-J model \cite{35,36,37,38,39,40,41}. For a detailed and rigorous comparison, we have considered the neutron scattering results from doped YBa$_2$Cu$_3$O$_{6+x}$ which has both Cu-O chains and planes and is a Mott-Hubbard insulator in the parental phase. Moreover, the doped phase can be very well described by the strongly correlated t-J model \cite{42}. The doping in YBa$_2$Cu$_3$O$_{6+x}$ introduces the holes only in the chains upto a critical doping concentration viz. x$_c\sim$0.41 and during this the valencies of copper and oxygen in the planes remain unchanged \cite{19}. Here it is important to point out that the randomly doped oxygen atoms in the chains convert the neighbouring Cu$^{+1}$ ions into Cu$^{+2}$ \cite{43}. The oxygen doping induces coupling between the spins situated on the chains, after a considerable amount of holes have been introduced  \cite{23}. The further doping introduces mobile holes and reduces the antiferromagnetic coupling between the itinerant spin degrees of freedom. As a consequence, the chains play a very important role in determining the response of the system to any externally applied perturbation, in the under-doped regime.\\
\hspace*{0.3cm} The wave vector dependent static spin susceptibility of YBa$_2$Cu$_3$O$_{6+x}$ can be extracted from the constant q-scans of the available neutron scattering data \cite{25}. The next step is to determine the effective spin exchange coupling J(q) for any stable magnetic state in an itinerant magnetic system. In general spin exchange constant can be shown to be directly related to the inverse of static wave-vector dependent spin susceptibility $\chi^{-1}$(q) \cite{27}. 
 To elaborate slightly on this, the non-local static magnetic susceptibility ($\tilde{\chi}$) can be expressed as the variation in spin magnetization `m' with respect to the external static magnetic field (H$_{ext}$) in the continuum case as follows \cite{27}:
\begin{align}
\tilde{\chi}(\overrightarrow{r},\overrightarrow{r\prime})=\frac{\delta m(\overrightarrow{r})}{\delta H_{ext}(\overrightarrow{r\prime})}=-\hspace*{0.2cm}\tilde{\chi}(\overrightarrow{r},\overrightarrow{r\prime})\hspace*{0.4cm}.\hspace*{0.4cm}\frac{\delta^2E}{\delta m(\overrightarrow{r})\delta m(\overrightarrow{r\prime})}\bigg{\vert}_{\mathsmaller{\lbrace{m(\overrightarrow{r})}\rbrace=\lbrace{m_r}\rbrace}}. \hspace*{0.1cm}\tilde{\chi}(\overrightarrow{r},\overrightarrow{r\prime})   
\end{align}
where $\lbrace$m$_r\rbrace$ denotes the magnetization at $\lbrace\overrightarrow{r}\rbrace$, representing any thermodynamically stable spin configuration and `E' is the ground state energy assuming the system to be at zero temperature. \\
Therefore from equation (23), we can write $\tilde{\chi}$ as \cite{27},
\begin{align}
\tilde{\chi}=\tilde{\chi}\hspace{0.2cm} \tilde{J}(\overrightarrow{r},\overrightarrow{r\prime})\hspace*{0.2cm}\tilde{\chi} 
\end{align}
where $\tilde{J}$($\overrightarrow{r},\overrightarrow{r\prime}$), the exchange coupling in real space connecting the spin magnetizations at $\overrightarrow{r}$ and $\overrightarrow{r\prime}$, is given by the usual definition \cite{27},
\begin{align}
\tilde{J}(\overrightarrow{r}\overrightarrow{r\prime})=
-\frac{\delta^2E}{\delta m(\overrightarrow{r})\delta m(\overrightarrow{r\prime})}\bigg{\vert}_{\mathsmaller{\lbrace{m(\overrightarrow{r})}\rbrace=\lbrace{m_r}\rbrace}}
\end{align}
The above equations are very general ones valid for any itinerant magnetic system. Now making use of equations (24) and (25) and for convenience calling $\tilde{J}$($\overrightarrow{r},\overrightarrow{r\prime}$)  as $\tilde{J}_{eff}$($\overrightarrow{r},\overrightarrow{r\prime}$) and $\tilde{\chi}$($\overrightarrow{r},\overrightarrow{r\prime}$) as $\tilde{\chi}_{eff}$($\overrightarrow{r},\overrightarrow{r\prime}$), appropriate to a doped quantum antiferromagnet (a special type of strongly correlated itinerant magnet), we arrive at the following relation,
\begin{align}
\tilde{J}_{eff}(\overrightarrow{r},\overrightarrow{r\prime})=\tilde{\chi}^{-1}_{eff}(\overrightarrow{r},\overrightarrow{r\prime})
\end{align}
Now, taking the Fourier transform,
\begin{align}
J_{eff}(q)=\chi_{eff}^{-1}(q)
\end{align}
It may be noted that the equation (27) is valid for all wave vectors, corresponding to any thermodynamically stable magnetic state, whether spontaneously magnetically ordered or not \cite{27,44}. Hence, this equation is valid even for the exchange coupled paramagnetic state with short-range spin correlations and may be used for determining J$_{eff}$(q) with the procedure being called `inverse susceptibility approach'.  \cite{27,44}. \\
Therefore, corresponding to the antiferromagnetic wave vector q=Q$\equiv\pi/a$,
\begin{subequations}
\begin{align}
J_{eff}(\pi/a)=\chi_{eff}^{-1}(\pi/a)
\end{align}
and for ferromagnetic coupling,
 \begin{align}
J_{eff}(0)=\chi_{eff}^{-1}(0)
\end{align}
\end{subequations}
\hspace*{0.3cm} Incidentally the above equations, valid even for a non-FL, look formally quite similar to the relation between the static spin susceptibility and the inverse of Landau parameter for a FL \cite{45}. \\
Using the minimization condition in equation(25), one can notice that all the above values of `J$_{eff}$' are negative. Here, we will concentrate on the variation of the absolute magnitude of `J$_{eff}$' with $\delta$, since the sign of `J' for ferromagnetic and antiferromagnetic coupling solely depends on the sign convention in writing the Hamiltonian (equation (2)).\\
Interestingly, the quantity D$_s$ represents another form of this spin-spin coupling in an itinerant magnet and its equivalence with J$_{eff}$ was established by us earlier \cite{12}. We now represent our results for the variation of D$_s^J$ and D$_s^t$ with $\delta$ in the 1D case. \\
\vspace*{6cm} 
\begin{subfigures}
\begin{figure}[!htb]
\minipage{0.5\textwidth}
{\includegraphics[width=\linewidth]{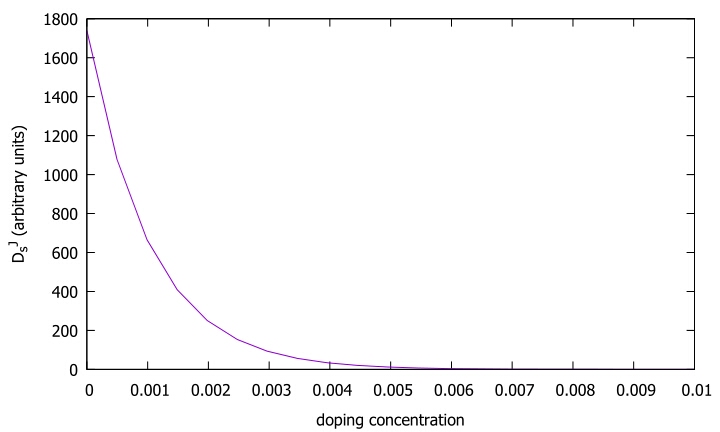}}\caption{\small{}}
\endminipage\hfill
\minipage{0.5\textwidth}
{\includegraphics[width=\linewidth]{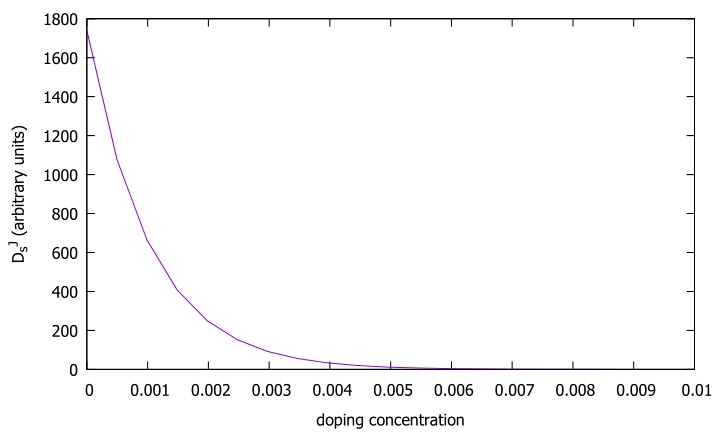}}\caption{\small{}}
\endminipage\hfill
\end{figure} 
\begin{figure}[!htb]
\centering
\includegraphics[scale=0.235]{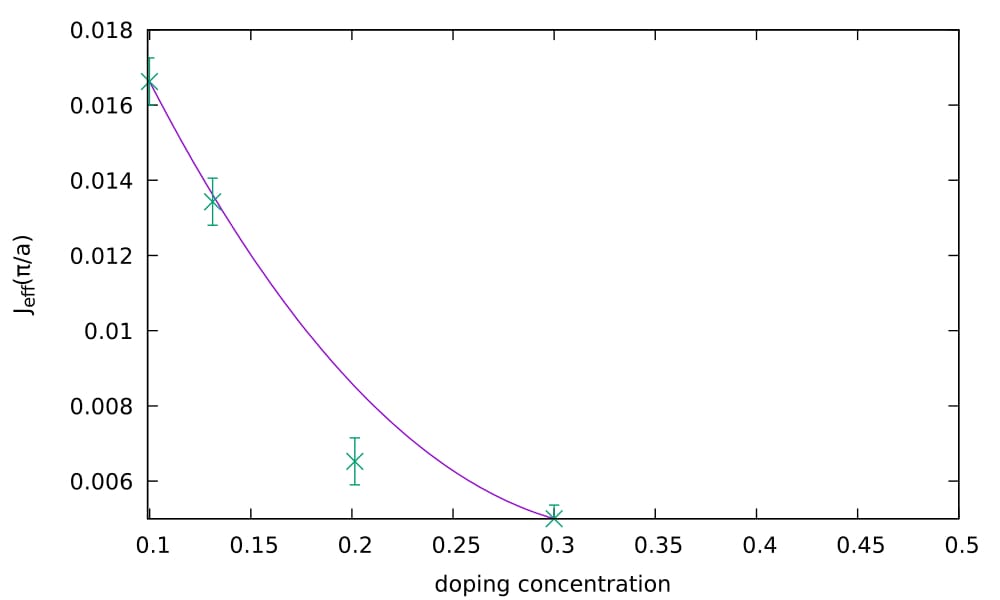} \caption{}
\end{figure}
\end{subfigures} \\ \\ \\

{\small{\textbf{Fig.(1)} `D$_s^J$' vs. doping concentration ($\delta$) for: {(a)} lattice length=1900; {(b)} lattice length=1960; {(c)} J$_{eff}$($\pi$/a) vs. $\delta$ obtained from the neutron scattering results on YBa$_2$Cu$_3$O$_{6+x}$. (In Fig.(c): The line gives the best polynomial fit to the data) \cite{28} }} \\ \\
\hspace*{0.3cm}The plots in (Figs.(1a,b)) show that D$_s^J$ falls drastically with increase in doping concentration and practically vanishes within $\delta\approx$0.005, which is much lower than the critical doping concentration observed for the 2D lattices \cite{12}. The suppression of D$_s^J$ corresponds to the fall in the ``semi-localized part'' of spin stiffness constant, which further implies the  destruction of original Heisenberg-like antiferromagnetic coupling in 1D with the introduction of very small amount of doping in the system. \\
 Fig.(1c) shows the fall of J$_{eff}$($\pi$/a) as a function of $\delta$, extracted from the constant q scans of neutron scattering data corresponding to q=Q$\equiv\pi$/a, which strongly resembles the behaviour our derived results of D$_s^J$ against $\delta$ \cite{28}. This rapid fall represents the loss in rigidity of the spins, as expected from the decay of semi-localized antiferromagnetism of the spins with Heisenberg-like character. 
\begin{subfigures}
\begin{figure}[!htb]
\minipage{0.5\textwidth}
{\includegraphics[width=\linewidth]{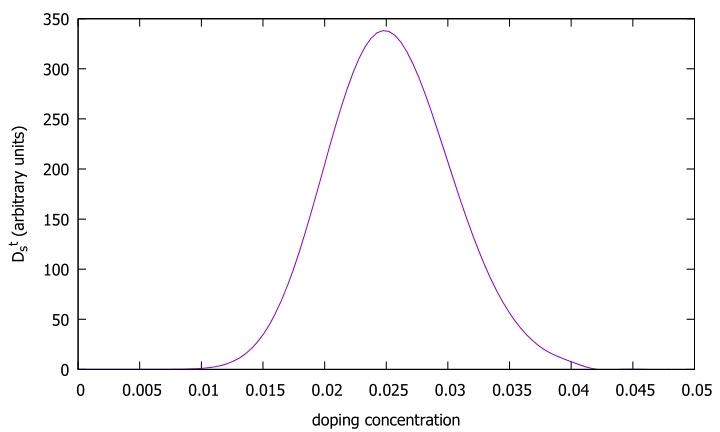}}\caption{\small{}}
\endminipage\hfill
\minipage{0.5\textwidth}
{\includegraphics[width=\linewidth]{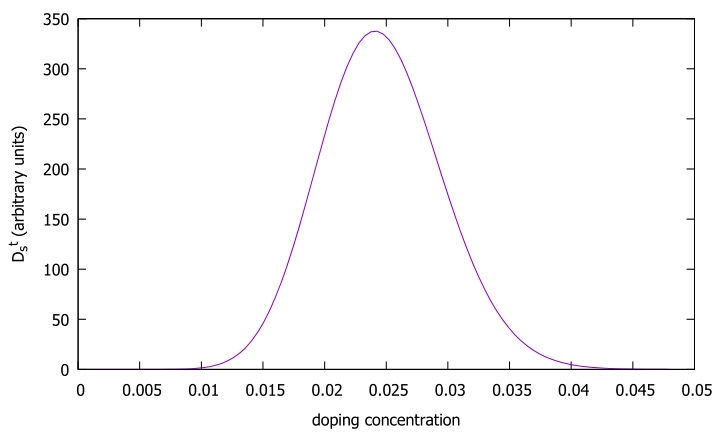}}\caption{\small{}}
\endminipage\hfill
\end{figure} 
\begin{figure}[H]
\centering
\includegraphics[scale=0.235]{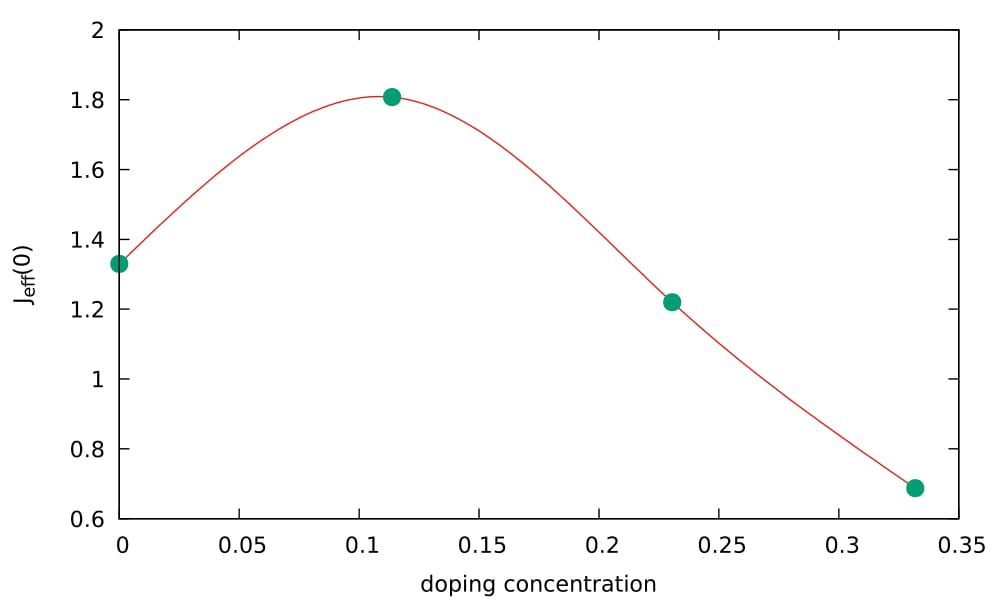} \caption{}
\end{figure}
\end{subfigures}
\vspace*{-0.1cm}
{\small{\textbf{Fig.(2)} `D$_s^t$' vs. doping concentration ($\delta$) for: {(a)} lattice length=1900; {(b)} lattice length=1960; {(c)} J$_{eff}$(0) vs. $\delta$ obtained from the dc susceptibilty measurements of YBa$_2$Cu$_3$O$_{6+x}$. (In Fig.(c): The line gives the best polynomial fit to the data) \cite{25} }} \\ \\
 Figs(2a,b) show that D$_s^t$ vanishes at $\delta$=0, which is also supported by our analytical results in the previous section. Beyond $\delta\rightarrow$0 limit, D$_s^t$ increases with $\delta$ in the very low doping regime, followed by a subsequent fall throughout the rest of the doping region. This characteristic behaviour leads to the appearance of a maximum in the D$_s^t$ vs. $\delta$ plot around 2$\%$-3$\%$ doping concentration. The increase in the magnitude of D$_s^t$ with $\delta$ and the appearance of the above peak signify the tendency of the itinerant spin degrees of freedom to develop another coupling, different from the original Heisenberg one. \\
\hspace*{0.3cm}The behaviour of experimentally extracted J$_{eff}$(0) with increase in $\delta$, in the low doping region, is shown in Fig.(2c) \cite{25}. J$_{eff}$(0) initially increases with increasing $\delta$ and again falls with further increase in doping concentration, giving rise to the appearance of a peak in the J$_{eff}$(0) vs. $\delta$ plot (see Fig.(2c)]. A maximum is also seen in our calculated D$_s^t$ vs. $\delta$ plot. Combining these experimental and theoretical results, we infer that there is a tendency of the itinerant spins in the system to develop a ferromagnetic-like coupling, corresponding to the wave vectors around q=0 [see Figs.(2a,b,c)). As $\delta$ is increased further, this coupling also gets greatly reduced and the spins become almost non-interacting to exhibit a behaviour analogous to Pauli-like in its magnetic response.  
\begin{figure}[H]
\centering
\includegraphics[scale=0.26]{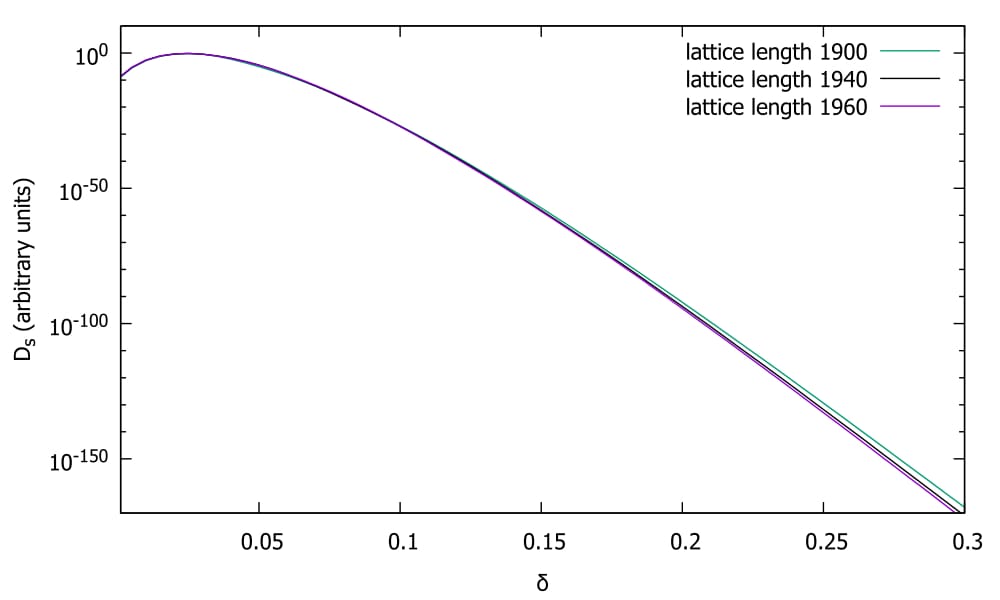} \caption{}
\end{figure}
\vspace*{-0.8cm}
\begin{center}
{\small{\textbf{Fig.(3)} Scaled spin stiffness constant (D$_s$) vs. doping concentration ($\delta$) for different lattice sizes in semi-logscale (Y-axis in log-scale)}}
\end{center}
\hspace{0.3cm}The Fig.(3) shows the plot of D$_s$ versus $\delta$ upto $\delta$=0.3. There exists a very sharp peak situated close to $\delta$=0, corresponding to the original quasi-long range antiferromagnetic ordering of semi-localized nature. However, this peak occurring at the extremely small region of doping concentration could not be shown in this figure. The presence of the peak at finite $\delta$ shown in Fig.(3), does represent an itinerant magnetic coupling tendency and this reminds one of the Stoner-like behaviour of the spin susceptibility, although in the strongly correlated background. Anyway, it must be pointed out that this itinerant coupling tendency does not form any long-range ordering in the system, even at zero temperature. The above plot shows that the slope of the fall of D$_s$ increases with increase in lattice size and the total spin stiffness practically vanishes much below 100$\%$ doping concentration. So it is worth mentioning that although analytical calculation proves the existence of rigidity of the spins upto 100$\%$ doping concentration, in reality the system becomes analogous to almost Pauli-paramagnetic-like at much lower value of doping concentration. \\
\hspace*{0.3cm}The comparison of our results with some other theoretical results shows agreements and certain disgreements, which highlight the crucial importance of our formalism in describing the quasi-1D doped antiferromagnetic systems and also clearly brings out the shortcomings of a few of those previous approaches. The singlet correlation between spins has been found previously using DMRG technique \cite{16}. The calculated spin correlation falls with the increase in doping concentration \cite{16}. As shown in Figs.(1-3), our derived D$_s^J$ rapidly decreases with $\delta$ and practically vanishes within the range of very low doping concentration. Thus, this is in agreement with DMRG results  in the low doping regime \cite{23}. Furthermore, the TMRG calculation shows an increase in static spin susceptibility ($\chi$(0)) with increase in $\delta$ \cite{17}. The increase in static uniform susceptibility results in the fall of J$_{eff}$(0) with $\delta$, as discussed in detail in the previous section (see equations(27-29)). So, this result is also in agreement with the behaviour of our calculated D$_s^t$ as a function of $\delta$, in the medium to over-doped regime.  \\ 

\section{Discussion} 
\hspace*{0.3cm} In the previous section, we have presented a detailed comparison of our calculational results with the relevant experimental and other theoretical ones, which firmly established that the generalized spin stiffness constants can play the role of effective exchange constants for quasi-1D doped quantum antiferromagnets as well, like that in the case of quasi-2D ones \cite{12}. Moreover, the results for 1D model are remarkably distinct from that of 2D and this striking contrast can have very different consequences for microscopic physics corresponding to various phenomena in quasi-1D and quasi-2D systems. \\ 
\hspace*{0.3cm} Our analytical and numerical results predicted the rapid decay of quasi-long-range ordered localized antiferromagnetic state, with doping, as expected from previous experimental results (see Fig.(1a,b)) \cite{26}. Most interestingly, our derived results in 1D, predicted the emergence of an unconventional itinerant paramagnetic phase with ferromagnetic spin-spin coupling, after the decay of quasi-long-ranged antiferromagnetically correlated one occurring at zero temperature. This novel prediction of ours on the magnetic behaviour of the system to display a tendency of short-ranged ferromagnetic-like ordering after the destruction of quasi-long range ordering in the 1D t-J model, should be taken up seriously by computational physicists and experimentalists for further investigation. \\ 
\hspace*{0.3cm}The prediction of the tendency of spin-reordering in 1D is in sharp contrast to our former results in 2D, which exhibited a possibility of a quantum phase transition in the over-doped regime \cite{12}. The presence of a point of quantum phase transition in 2D possibly denotes the existence of a separated hole-rich and hole-deficient phase below the critical value of doping concentration, as was previously predicted by numerical calculations \cite{6}. \\ 
\hspace*{0.3cm} The strongly correlated 2D models often show a transition from a non-Fermi liquid (NFL) strange metal phase to weakly correlated Fermi liquid (FL) phase in the mid-$\delta$ regime with the same value of U \cite{9,46}. In contrast, the t-J model shows the possibility of 1D fermionic systems for small J behaving as Tomonaga-Luttinger liquids (TLL) with the power law scaling of the correlation functions \cite{11,47}. This would be investigated by us in future. Though, in this paper we did not analyze the behaviour of the k-dependent momentum distribution function near k$_F$, nevertheless, we could well specify the magnetic phases and the corresponding phase boundaries throughout the range.\\ 
\hspace*{0.3cm} Our overall calculation has been done considering the simplified t-J model on a 1D tight binding lattice under a semi-continuum approximation. The lattice sizes taken are also much below the thermodynamic limit
(N$\rightarrow\infty$). The applicability of the strongly correlated t-J model to the over-doped regime in real materials is questionable, as the on-site Coulomb correlation between the holes in the system weakens drastically with increase in $\delta$ in the higher doping regimes. However, our non-perturbative calculations on the basis of the t-J model can correctly predict the magnetic behaviour of doped antiferromagnets keeping the `no double occupancy condition' intact, at least in the under-doped regime.\\
\hspace*{0.3cm} We have already stated in the preceding sections that our results are all derived considering only the nearest neighbour hopping and interactions between the spin degrees of freedom (t-J model). Regardless of this, the detailed comparisons of our results with the available experimental results, clearly demonstrates the success of our approach and formalism for analysing the magnetic correlations present in the low-dimensional strongly correlated doped antiferromagnetic systems. However, for real materials, the higher neighbour terms for both hopping and interaction play a significant role in determining the magnetic properties and even the high temperature superconducting phase boundaries of the hole-doped cuprates \cite{48}. Therefore, a deeper understanding of the magnetic phase boundaries of these materials requires the inclusion of higher neighbour hopping and interaction in the corresponding calculations. In this context, our recent work, considering the strongly correlated t$_1$-t$_2$-t$_3$-J model, shows that the point of possible quantum phase transition in 2D could be brought down to the mid-$\delta$ regime from the over-doped phase. This transition point incidentally may coincide with that of the cross-over  from anomalous Mott-Hubbard conducting phase to normal Fermi liquid-like metallic phase in real materials\cite{49,50,51,52}. Carrying out similar calculations for the 1D strongly correlated t$_1$-t$_2$-t$_3$-J model, we have found that the position of the peak in D$_s$ occurring at a finite $\delta$, gets shifted to lower values of $\delta$ and it reaches the $\delta\rightarrow$0 limit beyond the critical values of t$_2$ and t$_3$. The calculations including the higher neighbour exchange interactions are also necessary to strengthen our approach in future, for detailed investigation of the magnetic correlations in the doped quantum antiferromagnets. \cite{53,54,55}.\\      
\hspace*{0.3cm} To summarize, our calculations for generalized spin stiffness constant, introduces a comprehensive way for determining the evolution of the effective exchange coupling for the 1D strongly correlated t-J model with doping. With this, it is also possible to determine the magnetic phases appearing in the strongly correlated long-range antiferromagnetic insulators upon hole doping, in both one and two dimensions. The effective spin-spin interactions in the short-ranged ordered conducting paramagnetic phases showing NFL behaviour, have been studied with much rigour and precision. Further, the study of the interactions between the charge degrees of freedom, in terms of generalized charge stiffness constant (spin symmetric), to be taken up in near future with very similar formalism, would enable us to predict the complete phase boundaries of these low-dimensional systems more precisely. This would also provide clues to the possible pairing mechanism for superconductivity in the above materials.
 
\bibliographystyle{unsrt}

\end{document}